\documentclass[letterpaper, 10 pt, conference]{ieeeconf}  % Comment this line out if you need a4paper

\IEEEoverridecommandlockouts                              % This command is only needed if you want to use the \thanks command

\usepackage{graphics} % for pdf, bitmapped graphics files
\usepackage{epsfig} 
\usepackage{amsmath}
\usepackage{amssymb}
\usepackage{amsfonts}
\usepackage{stmaryrd} 
\usepackage{booktabs,tabularx, hhline,multicol}
\usepackage{algorithm,algpseudocode}
\usepackage{url}
\usepackage{breakurl}
\usepackage{epstopdf}
\usepackage{pbox}
\usepackage{bm}
\usepackage{enumerate}
\usepackage{mathtools}
\usepackage{caption}
\usepackage{subfigure}
\usepackage{bm}
\usepackage{float}
\usepackage{stfloats}
\usepackage{booktabs}
\usepackage{xcolor}
\usepackage{multirow}
\usepackage{balance}

\newtheorem{lem}{Lemma}

\newtheorem{defn}{Definition}
\newcommand{\enumbracket}[1]{{\llbracket #1 \rrbracket}}

\title{\LARGE \bf   Latent Dynamic Networked System Identification with High-Dimensional Networked  Data }

\author{ Jiaxin Yu,~\IEEEmembership{Student Member, IEEE}, Yanfang Mo,~\IEEEmembership{Member, IEEE}, and S. Joe Qin,~\IEEEmembership{Fellow, IEEE}
 \thanks{Jiaxin Yu is with the School of Data Science and Hong Kong Institute for Data Science,
City University of Hong  Kong, Hong Kong.}
\thanks{Yanfang Mo is with the School of Data Science and Hong Kong Institute for Data Science,
City University of Hong  Kong, Hong Kong. Co-corresponding author. {\tt\small Yanfang.MO@cityu.edu.hk}. }% <-this % stops a space
\thanks{S. Joe Qin is with the
Institute of Data Science, Lingnan University, Hong Kong. Co-corresponding author.
        {\tt\small sjoeqin@outlook.com}. He acknowledges the financial support from  a Natural Science Foundation of China Project (U20A20189), a General Research Fund by RGC of Hong Kong (No. 11303421),           
a Collaborative Research Fund by RGC of Hong Kong (Project No. C1143-20G), 
a grant from the Natural Science Foundation of China (U20A20189), 
        an ITF - Guangdong-Hong Kong Technology Cooperation Funding Scheme  (Project Ref. No. GHP/145/20),       
an InnoHK initiative of The Government of the
HKSAR for the Laboratory for AI-Powered Financial Technologies, 
and  a Shenzhen-Hong Kong-Macau Science and Technology Project Category C (9240086).}
}

\begin{document}

\maketitle
\thispagestyle{empty}
\pagestyle{empty}

\begin{abstract}                % Abstract of not more than 250 words.

Networked dynamic systems are ubiquitous in various domains, such as industrial processes, social networks, and biological systems. These systems produce high-dimensional data that reflect the complex interactions among the network nodes with rich  sensor measurements.  In this paper, we propose a novel algorithm for latent dynamic networked system identification that leverages  the network structure and performs dimension reduction for each node via dynamic  latent variables (DLVs). The algorithm assumes that the DLVs of each node have an auto-regressive model with exogenous input and  interactions from other nodes. The DLVs  of each node are extracted to capture the most predictable latent variables in the high dimensional data, while the residual factors are not predictable.  The advantage of the proposed framework is demonstrated on an industrial process network for system identification and dynamic data analytics.

\end{abstract}

\section{Introduction}
\label{sec:introduction}
% initial version, may add more contents about the networked models, and to be polished 
Networked dynamic systems have become increasingly prevalent across various fields, from industrial processes to social networks and biological systems. These systems are composed of interconnected nodes with their own dynamics and dynamic interactions with other nodes \cite{Wang:su:2014NetworkedSystems,zhou:etal:2018NetworkedSystems}.
The nodes in the networked systems are often equipped with  a rich set of sensors that yield high-dimensional time series data. Recent works have dealt with the estimation and system identification problems of these networked models with tremendous progress \cite{dankers:etal:2013predictors_in_networks,Zamani:etal:2015ID_networked,Haber:etal:2014SIM_interconnected}.
However, in dealing with high-dimensional networked systems data,  little has been done to extract low-dimensional latent dynamic networks from the high-dimensional networked data~\cite{Qin:etal:2020DRESS}.

To reduce dimension in data with static collinearity, statistical techniques such as principal component analysis (PCA), partial least squares (PLS), and canonical correlation analysis (CCA) are commonly used in data analytics~\cite{Kramer:Sugiyama:2011PLS_DoF,zhu:liu:qin:2017}. These methods effectively extract low-dimensional latent variables (LV) with statistical objectives. 
However, these methods do not focus on extracting latent dynamic information from data.

To deal with the issue of co-moving dynamics or collinear dynamics in high-dimensional data, 
dynamic latent variables (DLVs) and dynamic factor models (DFMs) have been developed.
Box and Tiao~\cite{box:tiao:1977} introduced a technique to decompose multivariate time series into dynamic and static latent factors. Following this work, DFMs have been extensively used in financial time series to extract reduced-dimensional dynamics~\cite{lam:yao:2012}. 
Besides, compressive sensing offers another perspective for extracting reduced-dimensional dynamic LVs (DLVs) with principles of signal processing and optimization~\cite{donoho:2009message}.

The aforementioned methods may not emphasize the predictability of the extracted DLVs. To this end, various reduced-dimensional DLVs methods with a focus on predictability have been proposed. For instance, the dynamic inner PCA (DiPCA)~\cite{dong:qin:2018DiPCA} algorithm, the dynamic inner PLS (DiPLS)~\cite{dong:qin:2018DiPLS} model, and the dynamic inner CCA (DiCCA)~\cite{dong:qin:2018DiCCA} method all extract the most predictable DLVs based on the univariate auto-regressive (AR) assumption. 
Recent works include the latent vector AR (LaVAR) model with fully interactive DLVs~\cite{Qin:2022LaVAR_AIChEJ} and the latent state space representation with a CCA objective (LaSS-CCA)~\cite{yu:qin:2022LaSS}. These methods also extract predictable DLVs in descending order and have been demonstrated to reduce the dimensionality of time-series data effectively. 
However, these methods do not consider the networked structure in the data, lacking the ability to capture the complex communications among the nodes of the networked system.

To address the challenges mentioned above, we propose a new framework for latent dynamic networked systems modeling with the following advantages: 
\begin{itemize}
    \item[1)] To handle co-moving or collinear dynamics, low dimensional dynamics are captured on each individual node system to extract DLVs with a network topology; 
    \item[2)] The proposed framework extends single-node DLV methods to networked dynamic systems to analyze high-dimensional networked time series data. 
    \item[3)] The proposed networked systems adopt vector auto-regressive models with exogenous input (VARX) for each node, and dynamic connections to other nodes follow a fully interactive topology; 
    \item[4)]  The identified network model is a  networked latent VARX model  (Net-LaVARX), which is readily suitable for networked system identification.
\end{itemize}
We demonstrate the effectiveness of the proposed  framework by applying it to an industrial process network. 

The remainder of the paper is organized as follows. In Section II, the proposed framework for latent dynamic  networked systems is presented. The Net-LaVARX algorithm is developed in Section III with its learning procedures. In Section IV, a dynamic case study is detailed to verify the proposed method. The conclusions are given in Section~V.

% \JQ{Notation defined by YF Mo }

\begin{table}[htpb]
\centering 
% \rev{
\caption{Notation}
\vspace{.5\baselineskip}
%\newcolumntype{Y}{>{\centering\arraybackslash}X}
  \label{tab:notation} 
  \begin{tabularx}{1\linewidth}{ll}
  \toprule    \toprule
 $\enumbracket{M}$ & node index set~$\{1,2,\ldots,M\}$\\
 $\mathcal{N}_i$& indices of incoming neighbor nodes, subset of   $\enumbracket{M}$ \\
 % $\mathcal{E}$& set of arcs in the network\\
 % $\mathcal{N}^O_i$ & out-neighbor set of Node~$i$, $\{j\mid (i,j) \in \mathcal{E}\}$ \\
%  $\bm e^i$ &the~$i$th column vector of the identity matrix~$\mathbf{I}$\\
  %$\mathbf{A}=(a_{ij})$& adjacency matrix of the network\\
  $\bm y^i_k$ & measurement vector of Node~$i$\\
  $\bm u^i_k$ & exogenous input of Node~$i$\\
  $\bm v^i_k$ & dynamic latent variables (DLVs) of Node~$i$\\
 {$\bm{\varepsilon}^i_k$ } & innovations vector of Node~$i$, driving $\bm v^i_k$\\
  $p_i/\ell_i/m_i$ & dimension of the measurement/DLVs/input of Node~$i$\\
  $s_i$ & DLV order of Node~$i$\\
  $\mathbf{R}_i$ & projection weight matrix of Node~$i$\\
  $\mathbf{P}_i$ & DLV loadings matrix of Node~$i$  \\
  $\mathbf{W}_i$ & equivalent projection weight matrix of Node~$i$ \\
  $\mathbf{\bar P}_i$ & static loadings matrix of Node~$i$  \\
  %$\bm {\varepsilon}^{O_i}_t/\bm {\varepsilon}^{I_i}_t$ & outer/inner model noise on Node~$i$\\
  % $s_i$ & order of the input model on Node~$i$\\
  %$s_{ii}$ & order of the intra X part in Node~$i$\\
  %$s_{ij}$ & order of dynamics in Arc~$(i,j)\in \mathcal{E}$;\\
  $\mathbf{A}^{i}_h$ & auto-regressive coefficient matrix of Node  $i$ \\
  $\mathbf{B}^{i}_h$ & exogenous input coefficient matrix on Node~$i$\\
  $\mathbf{C}^{ij}_h$ & regression coefficient matrix from Node $j$ to Node $i$ \\
  %$\mathbf{A}^{ii}_k$ & autoregressive coefficient matrices on Node~$i$\\
  %$(i,j)$ & arc with tail node~$i$ and head node~$j$\\
  \bottomrule \bottomrule
  \end{tabularx} %}
\end{table}

\section{ Networked Latent Dynamic   Systems}

\label{secII}

\subsection{Networked latent vector AR model with exogenous input (Net-LaVARX)}
% \rev{
As described in the introduction section, high dimensional data usually have dynamics with a reduced dimension for each node of a networked system, giving rise to the need to develop networked latent dynamic modeling methods. For convenience, the notation is listed in Table~\ref{tab:notation}.
Consider a networked dynamic system of~$M$ nodes, where Node $i\in\enumbracket{M}$ has a measurement vector~$ \bm{y}^{i}_{k} \in \mathbb{R}^{p_i}$.
Let~$ \bm{v}^{i}_{k} \in \mathbb{R}^{\ell_i}$ represent the reduced dimensional dynamic latent vector (DLV) for Node $i$, whose elements lie independently in  an~$\ell_i$-dimensional subspace. Following \cite{Qin:2022LaVAR_AIChEJ} for a single node system, the dynamics  of $\bm{y}^{i}_{k}$ are related to $\bm{v}^{i}_{k}$ with an  outer model,
\begin{align} \label{eq:NLaVARX_outer}
    \bm y^i_k = \mathbf{P}_i \bm v^i_k
    +  \mathbf{\bar P}_i \bm{\bar \varepsilon} ^i_k,
\end{align}
where~$\bm{\bar \varepsilon}^i_k\in \Re^{p_i - {\ell_i}}$ denotes the static noise, and~$\mathbf{P}_i \in \Re^{p_i\times \ell_i}$ and $\mathbf{\bar P}_i \in \Re^{p_i\times (p_i - {\ell_i})}$ are the respective loadings for the DLVs and static latent factors. 

It is noted that the auto-correlated  DLVs $\bm{v}^{i}_{k}$ and static noise 
$\bm{\bar \varepsilon}^i_k$ complement each other to make up the whole signal series $\bm{y}^{i}_{k}$. We give the following definition for a networked latent vector auto-regressive model with exogenous input.

\begin{defn} \label{def:NetLaVARX}
A networked latent vector auto-regressive model with exogenous input (Net-LaVARX) system is described by \eqref{eq:NLaVARX_outer} with 
\begin{enumerate}
    \item $\left[\mathbf{P}_i \quad 
    \mathbf{\bar  P}_i  \right] \in \Re^{p_i\times p_i} $ form a non-singular matrix,
    \item the DLVs $\bm v^i_k$ of  Node~$i$ have auto-regressive terms with exogenous input  as
\begin{equation}  \label{eq:netLaVARX}
    \hspace{-2em}\bm v^i_k = 
     \sum_{h=1}^{s_i} \left( \mathbf{A}^{i}_{h}\bm v^i_{k-h} +  \mathbf{B}^{i}_{h}\bm u^i_{k-h} \right) +
     \sum_{j \in \mathcal{N}_i}\sum_{h=1}^{s_i} 
     \mathbf{C}^{ij}_{h}\bm v^j_{k-h} 
    + \bm{\varepsilon}^i_k,
\end{equation}
% \begin{equation}  \label{eq:netLaVARX}
%     \hspace{-2em}\bm v^i_k \!=\! 
%      \sum_{h=1}^{s_i}\! \left(\! \mathbf{A}^{i}_{h}\bm v^i_{k-h} +  \mathbf{B}^{i}_{h}\bm u^i_{k-h}  + \!\!
%      \sum_{j \in \mathcal{N}_i}\!
%      \mathbf{C}^{ij}_{h}\bm v^j_{k-h} \!\right)
%     + \bm{\varepsilon}^i_k,
% \end{equation}
where
the notation is given in Table \ref{tab:notation}, and 
\item {the noise terms $\bm{\varepsilon} ^i_k, ~ \bm{\bar \varepsilon}^j_k$ are serially uncorrelated and mutually independent random sequences for $i\in\enumbracket{M}$.} 
\end{enumerate}
% \JQ{ $[\bm{\varepsilon} ^i_k \quad \bm{\bar \varepsilon} ^j_k]$ often means an augmented matrix; this expression is not good since the dimensions are not compatible.  }
\end{defn}

\begin{figure}[t]
    \centering
\includegraphics[width=0.92 \columnwidth]{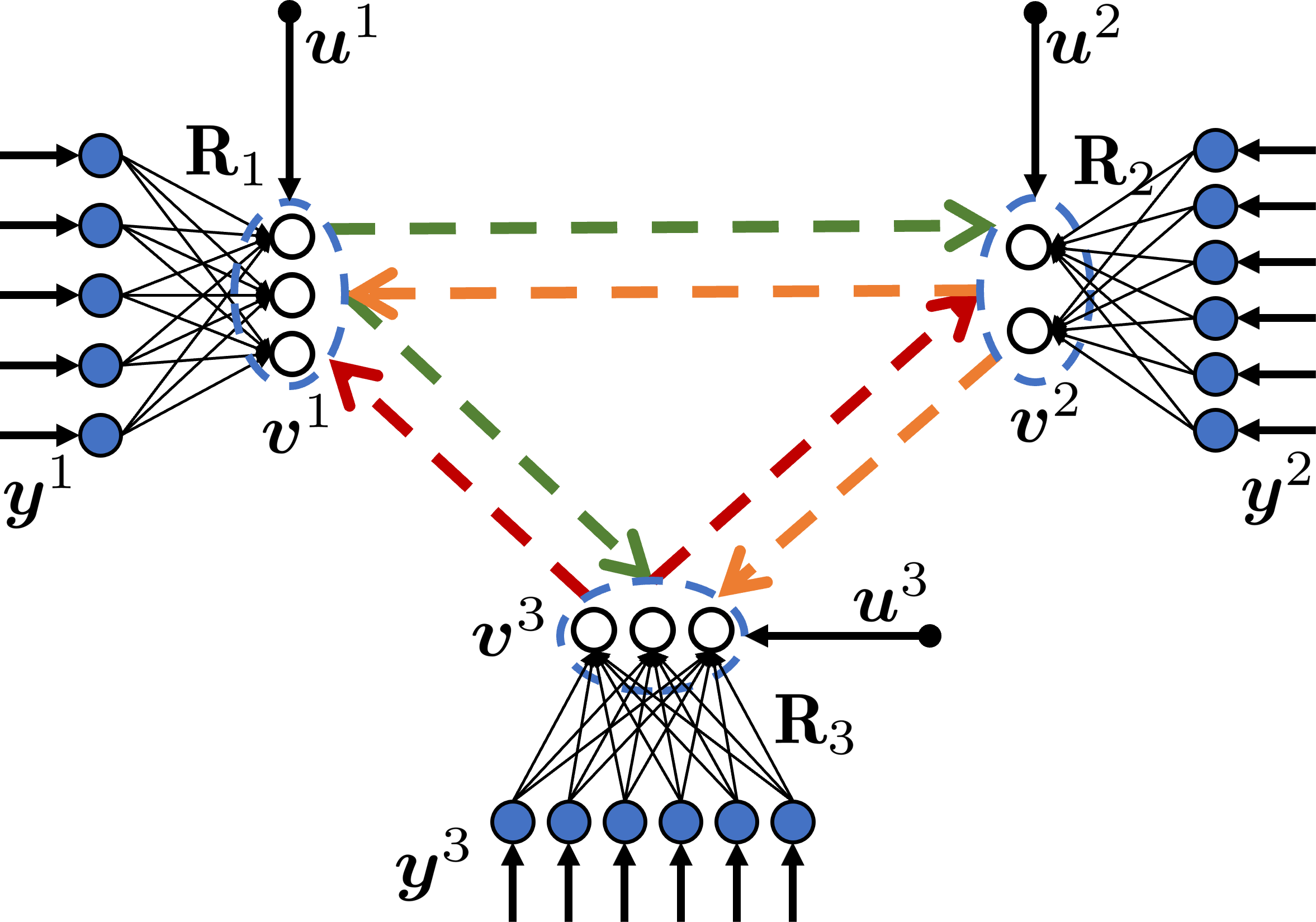}
    \caption{A diagram of a networked latent dynamic systems with three nodes.}% \rev{and five arcs}
    \label{fig:diagram}
\end{figure}

With the definition,  if $\ell_i = p_i$, the model is a standard full-dimensional networked dynamic system, and $\bm{\bar \varepsilon} ^i_k$ is null. On the other hand, if $\ell_i = 0$, the model is composed of static nodes only. Without the cross terms $\mathbf{C}^{ij}_{h}$, the above model is a standard latent VARX model for Node $i$. 
The networked dynamic system is illustrated in Fig.~\ref{fig:diagram}, where dashed arrows represent dynamic interactions between nodes.

\subsection{Extracting Net-LaVARX DLVs via oblique projections }

The ultimate objective of the Net-LaVARX modeling is to develop algorithms to extract the DLVs $\bm v^i_k$ and estimate model parameters of \eqref{eq:NLaVARX_outer} and \eqref{eq:netLaVARX}. 
To do this, we design a projection to eliminate the column space of $\mathbf{\bar  P}_i$ and focus on that of $\mathbf{  P}_i$. We give the following lemma to state the result. 

\begin{lem} \label{lem:proj_matrix}
Let $\mathbf{\bar  P}_i^\perp \in \Re^{p_i \times \ell_i} $ have full column rank and  $\mathbf{\bar  P}_i^\intercal \mathbf{\bar  P}_i^\perp =\mathbf{0} $. Then 
\begin{align} \label{eq:NLaVARX_vk}
 \bm v^i_k = \mathbf{R}_i^{\intercal} \bm y^i_k,
\end{align}
where $\mathbf{R}_i = \mathbf{\bar  P}_i^\perp
(\mathbf{  P}_i^\intercal  \mathbf{\bar  P}_i^\perp)^{-1} $, $\mathbf{R}_i ^\intercal \mathbf{P}_i = \mathbf{I}$, and 
$\mathbf{P}_i \mathbf{R}_i ^\intercal$ is an oblique projection matrix, i.e.,~$(\mathbf{P}_i \mathbf{R}_i ^\intercal)^2 = 
\mathbf{P}_i \mathbf{R}_i ^\intercal $.  

%%% original version
%Letting $\mathbf{\bar  P}_i^\perp \in \Re^{p_i \times \ell_i} $ be an orthogonal complement of $\mathbf{\bar  P}_i$, which
%has full column rank and 
%$\mathbf{\bar  P}_i^\intercal \mathbf{\bar  P}_i^\perp =\mathbf{0} $, we have 
% \begin{align} \label{eq:NLaVARX_vk}
%  \bm v^i_k = \mathbf{R}_i^{\intercal} \bm y^i_k,
% \end{align}
% where $\mathbf{R}_i = \mathbf{\bar  P}_i^\perp
% (\mathbf{  P}_i^\intercal  \mathbf{\bar  P}_i^\perp)^{-1} $ and 
% $\mathbf{P}_i \mathbf{R}_i ^\intercal$ is an oblique projection matrix which satisfies 
% $(\mathbf{P}_i \mathbf{R}_i ^\intercal)^2 = 
% \mathbf{P}_i \mathbf{R}_i ^\intercal $
% and
% $\mathbf{R}_i ^\intercal \mathbf{P}_i = \mathbf{I}$.
\end{lem}

The proof of this lemma is given in the appendix of the paper. This lemma shows that we can extract the DLVs via an oblique projection. Further, using \eqref{eq:NLaVARX_vk}, the projection 
\begin{align} \label{eq:NLaVARX_proj }
\bm y^{i,\texttt{proj}}_k = 
\mathbf{P}_i \mathbf{R}_i^{\intercal} \bm y^i_k = \mathbf{P}_i \bm v^i_k 
\end{align}
is a reconstruction of $\bm y^i_k$. 

While separating the dynamic variations from static ones in \eqref{eq:NLaVARX_outer} is feasible, $\mathbf{P}_i$ cannot be uniquely identified due to the bilinear decomposition in \eqref{eq:NLaVARX_outer}. Assuming  $\bm v_k^{*i}$ is the true DLV with loadings $\mathbf{P}_i^*$, we have 
\[
\bm y^i_k = \mathbf{P}_i^* \bm v_k^{*i}
    +  \mathbf{\bar P}_i \bm{\bar \varepsilon} ^i_k.
\]
For any nonsingular $\mathbf{M} \in \Re^{\ell_{i} \times \ell_{i}} $, we have 
\[
\mathbf{P}_i^* \bm v_k^{*i} = (\mathbf{P}_i^* \mathbf{M}^{-1})( \mathbf{M} \bm v_k^{*i}) 
\]
It is obvious that 
$ \bm v_k^{i} = \mathbf{M} \bm v_k^{*i}$ is an equivalent dynamic latent vector. Therefore, the true  loadings matrix can only be identified up to the same range space, since $\mathbf{P}_i^* \mathbf{M}^{-1}$ and 
$\mathbf{P}_i^*$ share the same range space. 
For this reason, the subspace of the DLVs is uniquely identified.  There  are extra degrees of freedom to make the DLVs in $\bm v_k^{i}$ in a pre-specified order, e.g., a descending order of predictability with  constrained magnitudes \cite{dong:qin:2018DiCCA,Qin:2022LaVAR_AIChEJ}.

Substituting \eqref{eq:NLaVARX_vk}, \eqref{eq:netLaVARX} into \eqref{eq:NLaVARX_outer} gives 
the model of measurement variables as
\begin{align} \label{eq:NLaVARX_yk}
     \bm y^i_k&=  \sum_{h=1}^{s_i} ( \mathbf{P}_i 
 \mathbf{A}^{i}_{h} \mathbf{R}_i^\intercal \bm  y^i_{k-h} + \mathbf{P}_i \mathbf{B}^{i}_{h}\bm u^i_{k-h}) \nonumber \\
     & + 
     \sum_{j \in \mathcal{N}_i} \sum_{h=1}^{s_i} 
    \mathbf{P}_i \mathbf{C}^{ij}_{h} \mathbf{R}_i^\intercal \bm  y^j_{k-h}  + \bm e^i_k, 
\end{align}
which is a networked VARX model with reduced-rank auto-regressive matrices 
$\mathbf{P}_i 
 \mathbf{A}^{i}_{h} \mathbf{R}_i^\intercal$, 
where 
\[
\bm e^i_k = \mathbf{P}_i \bm \varepsilon^i_k + 
\mathbf{\bar P}_i \bm {\bar \varepsilon}^i_k
\]
is the equivalent noise sequence of the model.

Since $\bm \varepsilon^i_k \in \Re^{{\ell}_{i}}$ and $\bm{e}_{k}^{{i}} \in \Re^{p_{i}}$ are the random noises that are serially independent, the best predictions of ${\bm{v}}^{i}_{k}$ and ${\bm{y}}^{i}_{k}$ are, respectively
\begin{align} \label{eq:vk_hat}
 \bm{\hat v}^i_k&=  \sum_{h=1}^{s_i} ( 
 \mathbf{A}^{i}_{h}  \bm  v^i_{k-h} +  \mathbf{B}^{i}_{h}\bm u^i_{k-h} )  
 + \sum_{j \in \mathcal{N}_i} \sum_{h=1}^{s_i} 
  \mathbf{C}^{ij}_{h}  \bm  v^j_{k-h},  \\
  \bm{\hat y}^i_k&=  \mathbf{P}_{i}
 \bm{\hat v}^i_k
\end{align}
with the auto-regressive coefficient matrix $\mathbf{A}^{i}_{h} \in \Re^{\ell_{i} \times \ell_{i}}$, the exogenous input coefficient matrix $\mathbf{B}^{i}_{h} \in \Re^{\ell_{i} \times m_{i}}$, and the regression coefficient matrix $\mathbf{C}^{ij}_{h} \in \Re^{\ell_{i} \times \ell_{j}}$.

\subsection{Net-LaVARX with a CCA objective }

Taking account of the interactions between the networked nodes, we aim to find the projection weight matrices $\mathbf{R}_{i}$ for~$i\in \enumbracket{M}$ that make the DLVs $\bm{v}^{i}_{k}$ the most predictable based on the latent dynamics, with {$\ell_{i} <  p_{i}, \forall i\in \enumbracket{M}$} 
% Yes, it needs to be \ell_{i} <  p_{i} to be consistent with the optimal num of DLV in Table III.
for the dimension reduction purposes. Let $s = \max \{s_i,i\in \enumbracket{M}\} $
and consider that we have $s+N$ samples of training data $\{ \{ \bm{y}^{i}_{k}, \bm{u}^{i}_{k} \}_{k=1}^{s+N},  i \in \enumbracket{M} \}$. 
Similar to \cite{Qin:2022LaVAR_AIChEJ} for a single node system, we use the last $N$ samples to form 
the following Net-LaVARX CCA objective 
\begin{align} \label{eq:CCA_obj_sum}
   \min_{ \{ \mathbf{R}_i, \mathbf{A}^{i}_{h}, \mathbf{B}^{i}_{h}, \mathbf{C}^{ij}_{h} \} } & \ \  
   J = \sum_{i=1}^M   \sum_{k=s+1}^{s+N} \|\bm{v}^{i}_{k} -   \bm{\hat v}^{i}_{k}\|^2 \\
   \texttt{s.t.} & \quad \sum_{k=s+1}^{s+N} \bm{v}^{i}_{k} (\bm{v}^{i}_{k})^\intercal = \mathbf{I}, \quad i \in \enumbracket{M},
\end{align}
where the first $s$ samples are saved for initialization.

\section{Learning Procedure of Dynamic Latent Network Systems}

\subsection{Net-LaVARX objective with training data }

With $s+N$ samples of training data $\{ \{ \bm{y}^{i}_{k}, \bm{u}^{i}_{k} \}_{k=1}^{s+N},  i \in \enumbracket{M} \}$, we denote the matrices of output and input data as
\begin{equation}
\begin{aligned}
\mathbf{Y}^{i} &= \left[ \begin{array}{llll}
\bm{y}^{i}_{1} & \bm{y}^{i}_{2} & \cdots & \bm{y}^{i}_{s+N}
\end{array} \right]^{\intercal} \in \Re^{(s+N) \times p_{i}},  \\
\mathbf{U}^{i} &= \left[ \begin{array}{llll}
\bm{u}^{i}_{1} & \bm{u}^{i}_{2} & \cdots & \bm{u}^{i}_{s+N}
\end{array} \right]^{\intercal} \in \Re^{(s+N) \times m_{i}}.
\end{aligned}
\end{equation}
The time-shifted matrices with $N$ samples are formed as
\begin{equation}
\begin{aligned}
\mathbf{Y}^{i}_{j} &=\left[ \begin{array}{llll}
\bm{y}^{i}_{j+1} & \bm{y}^{i}_{j+2} & \cdots & \bm{y}^{i}_{j+N}
\end{array} \right]^{\intercal} \in \Re^{N \times p_{i}}\\
\mathbf{U}^{i}_{j} &= \left[\begin{array}{llll}
\bm{u}^{i}_{j+1} & \bm{u}^{i}_{j+2} & \cdots & \bm{u}^{i}_{j+N}
\end{array}\right]^{\intercal} \in \Re^{N \times m_{i}} \\
\mathbf{V}^{i}_{j} &=\left[\begin{array}{llll}
\bm{v}^{i}_{j+1} & \bm{v}^{i}_{j+2} & \cdots & \bm{v}^{i}_{j+N}
\end{array}\right]^{\intercal} \in \Re^{N \times \ell_{i}}, \\
\end{aligned}
\end{equation}
for $ j=0,1, \ldots, s_{}$, where $\mathbf{V}^{i}_{j} =\mathbf{Y}^{i}_{j} \mathbf{R}_{i}$.
For the case of 
$j=s_{}$, we have
\[
\mathbf{V}^{i}_s=\left[\begin{array}{llll}\bm{v}^{i}_{s_{}+1} & \bm{v}^{i}_{s_{}+2} & \cdots & \bm{v}^{i}_{s_{}+N}\end{array}\right]^{\intercal}
\in \Re^{N \times \ell_{i}}
\]
The objective \eqref{eq:CCA_obj_sum} is reformulated as 
\begin{align} \label{eq:CCA_obj}
   \min_{ \{ \mathbf{R}_i, \mathbf{A}^{i}_{h}, \mathbf{B}^{i}_{h}, \mathbf{C}^{ij}_{h} \} } & \ \  
   J = \sum_{i=1}^M   \|\mathbf{V}^{i}_s - \mathbf{\hat V}^{i}_s\|_F^2 \\
   \texttt{s.t.} & \quad (\mathbf{V}^{i}_s)^\intercal
   \mathbf{V}^{i}_s = \mathbf{I}, \quad i \in \enumbracket{M}
\end{align}
where  $\|\cdot\|_{F}$ denotes the Frobenius norm and $\mathbf{\hat V}^{i}_s$ depends on 
$\{ \mathbf{R}_i, \mathbf{A}^{i}_{h}, \mathbf{B}^{i}_{h}, \mathbf{C}^{ij}_{h} \}$ which can be obtained from \eqref{eq:vk_hat} as
\begin{align}
\mathbf{\hat V}^{i}_s
&= \sum_{h=1}^{s_i} \left(
\mathbf{V}^{i}_{s-h}
{\mathbf{A}^{i}_{h}}^{\intercal}   + \mathbf{U}^i_{s-h} 
{\mathbf{B}^{i}_{h}}^{\intercal} \right) + \sum_{j \in \mathcal{N}_i} \sum_{h=1}^{s_i} \mathbf{V}^{j}_{s-h} {\mathbf{C}^{ij}_{h}}^{\intercal} \nonumber \\
    &= \mathbb{V}^{i}_s \mathbb{A}^{i} +
    \mathbb{U}_s^{i} {\mathbb{B}}^{i} + 
    \sum_{j \in \mathcal{N}_i} \mathbb{V}^{j}_s \mathbb{C}^{ij},
\end{align}
where the augmented matrices are given as follows 
\begin{align*}
{\mathbb{V}}_{s}^{j} &=\left[\begin{array}{llll}
\mathbf{V}^{j}_{s-1} & \mathbf{V}^{j}_{s-2} & \cdots & \mathbf{V}^{j}_{s-s_{i}}
\end{array}\right], \quad j \in \enumbracket{M} \\
\mathbb{U}_s^{i} &=\left[\begin{array}{llll}
\mathbf{U}^{i}_{s-1} & \mathbf{U}^{i}_{s-2} & \cdots & \mathbf{U}^{i}_{s-s_{i}}
\end{array}\right], \\
{\mathbb{A}}^{i} &=\left[\begin{array}{llll}
{\mathbf{A}^{i}_{1}} & {\mathbf{A}^{i}_{2}} & \cdots & {\mathbf{A}^{i}_{s_i}}
\end{array}\right]^{\intercal},\\
{\mathbb{B}}^{i} &=\left[\begin{array}{llll}
{\mathbf{B}^{i}_{1}} & {\mathbf{B}^{i}_{2}} & \cdots & {\mathbf{B}^{i}_{s_i}}
\end{array}\right]^{\intercal},\\
{\mathbb{C}}^{ij} &=\left[\begin{array}{llll}
\mathbf{C}^{ij}_{1} & \mathbf{C}^{ij}_{2} & \cdots & \mathbf{C}^{ij}_{s_i}
\end{array}\right]^{\intercal}, \quad j \in \mathcal{N}_i
\end{align*}
for $i\in \enumbracket{M}$. To further simplify the notations, we {denote} % the concatenation of DLVs and inputs augmentation as
 \[
 \mathbb{Z}^{i} =  \begin{bmatrix}
        \mathbb{V}^{i} &  \mathbb{U}^{i} &
        [\mathbb{V}^{j}]_{j\in \mathcal{N}_i} 
        \end{bmatrix},
 \]
where $[\mathbb{V}^{j}]_{j\in \mathcal{N}_i} $  is arranged row-wise with all $\mathbb{V}^{j}$ for $j\in \mathcal{N}_i$. Similarly, we {collect all DLV model parameters as} 
% \rev{ \mathbb{Q} should be the DLV model parameters, or call it VARX model parameters?} 
 \[
 \mathbb{Q}^{i} =  \begin{bmatrix}
        \mathbb{A}^{i} \\  \mathbb{B}^{i} \\
        [\mathbb{C}^{ij}]_{j\in \mathcal{N}_i} 
        \end{bmatrix},
 \]
 where $[\mathbb{C}^{ij}]_{j\in \mathcal{N}_i} $  is arranged column-wise with all $\mathbb{C}^{ij}$ for $j\in \mathcal{N}_i$.
 The objective \eqref{eq:CCA_obj} is rewritten  as 
\begin{align} \label{eq:CCA_obj_RQ}
   \min_{ \{ \mathbf{R}_i, \mathbb{Q}^{i} \} } & \ \  
   J = \sum_{i=1}^M   \|\mathbf{V}^{i}_s - \mathbf{\hat V}^{i}_s\|_F^2 = \sum_{i=1}^M   \|\mathbf{V}^{i}_s -  \mathbb{Z}^{i} \mathbb{Q}^{i} \|_F^2 \nonumber \\
   \texttt{s.t.} & \quad (\mathbf{V}^{i}_s)^\intercal
   \mathbf{V}^{i}_s = \mathbf{I}, \quad i \in \enumbracket{M}.
\end{align}
 The above equality constraints guarantee that the objective is equivalent to the sum of canonical correlations of all nodes \cite{Qin:2022LaVAR_AIChEJ}.

\subsection{Solution of the Net-LaVARX models}

Due to the bilinear decomposition of the Net-LaVARX models, an alternating iterative solution is expected, as shown in early work~\cite{Qin:2022LaVAR_AIChEJ}.
Below we give an iterative solution to update $\{ \mathbf{R}_{i} \}_{i =1}^M $. 
Given $\{ \mathbf{R}_{i} \}_{i =1}^M $, $\{ \mathbb{V}_s^{i} \}_{i =1}^M $ can be calculated.  
We have the following solution for the $i$-th component by minimizing~\eqref{eq:CCA_obj_RQ},
\begin{equation}
\label{eq:sol_Q_given_R}
    \mathbb{Q}^{i} = 
    {(\mathbb{Z}^{i})}^{+} \mathbf{V}^{i}_s,
\end{equation}
{where  $(\cdot)^{+}$ refers to the Moore–Penrose pseudo-inverse}. Then we have 
\begin{equation}
    \hat{\mathbf{V}}^{i}_s = \mathbb{Z}^{i} {(\mathbb{Z}^{i})}^{+} \mathbf{V}^{i}_s.
\end{equation}
The objective~\eqref{eq:CCA_obj_RQ} can be rewritten as
\begin{equation}
\begin{aligned}
\label{eq:J_K=M}
\min_{\{ \mathbf{R}_i \}} & \ \  J_{} 
= \sum_{i=1}^{M} \left\|\mathbf{V}^{i}_s - 
        \mathbb{Z}^{i} {(\mathbb{Z}^{i})}^{+} \mathbf{V}^{i}_s  \right\|_{F}^{2} \\ 
    & \ \ \ \ = \sum_{i=1}^{M} \left\| \mathbf{V}^{i}_s \right\|_{F}^{2} - \sum_{i=1}^{M} \left\|  \mathbb{Z}^{i} {(\mathbb{Z}^{i})}^{+} \mathbf{V}^{i}_s \right\|_{F}^{2} \\
    & \ \ \ \ = \sum_{i=1}^{M} \ell_{i}  - \sum_{i=1}^{M} \left\|  \mathbb{Z}^{i} {(\mathbb{Z}^{i})}^{+} \mathbf{Y}^{i}_s \mathbf{R}_{i} \right\|_{F}^{2} \\
    \texttt {s.t.} & \quad (\mathbf{V}^{i}_{s_{}})^{\intercal} \mathbf{V}^{i}_{s_{}} = \mathbf{I}_{\ell_{i}},  \quad i \in \enumbracket{M}
\end{aligned}
\end{equation}
In the iterative scheme, $\mathbb{Z}^{i}$ depends on the previous values of $\{ \mathbf{R}_{i} \}_{i=1}^M$. Therefore, 
the above objective is equivalent to maximizing the following  $M$ separate objectives, 
\begin{equation} \label{eq:CCA_obj_Ji}
\begin{aligned}
\max_{\mathbf{R}_{i}} \ \  J_i  & = \left\|  \mathbb{Z}^{i} {(\mathbb{Z}^{i})}^{+} \mathbf{Y}^{i}_s\mathbf{R}_{i} \right\|_{F}^{2} \\
    \texttt {s.t. } & \quad {(\mathbf{V}^{i}_s)}^{\intercal} \mathbf{V}^{i}_s = \mathbf{R}_{i}^{\intercal}                {(\mathbf{Y}^{i}_s)}^{\intercal}  \mathbf{Y}^{i}_s \mathbf{R}_{i} = \mathbf{I}_{\ell_{i}}
\end{aligned}
\end{equation}
for $i \in \enumbracket{M}$. 
To make it easy to satisfy the equality constraints, we perform the following singular value decomposition (SVD) 
\[
\mathbf{Y}_{s}^{i}  = \mathbf{\cal U}_{s}^{i} \mathbf{\cal D}_{s}^{i} {(\mathbf{\cal V}_{s}^{i})}^{\intercal}
\]
where $\mathbf{\cal D}_{s}^{i}$ retains all non-zero singular values of $\mathbf{Y}_{s}^{i}$ which makes $ \mathbf{\cal U}_{s}^{i} = \mathbf{Y}_{s}^{i} \mathbf{\cal V}_{s}^{i} (\mathbf{\cal D}_{s}^{i} )^{-1}$. The DLV score matrix is equivalently obtained by linearly combining $\mathbf{\cal U}_{s}^{i}$ as
\[
\mathbf{V}^{i}_s = \mathbf{\cal U}_{s}^{i} \mathbf{W}_{i}
= \mathbf{Y}_{s}^{i} \mathbf{\cal V}_{s}^{i} (\mathbf{\cal D}_{s}^{i} )^{-1} \mathbf{W}_{i}
= \mathbf{Y}^{i}_s \mathbf{R}_{i}
\]
which makes 
\begin{align} \label{eq:Ri_from_Wi}
 \mathbf{R}_{i} = \mathbf{\cal V}_{s}^{i} (\mathbf{\cal D}_{s}^{i} )^{-1} \mathbf{W}_{i} 
\end{align}
Therefore, solving $\mathbf{W}_{i}$ is equivalent to solving for $\mathbf{R}_{i}$.

With the above SVD the objective \eqref{eq:CCA_obj_Ji} is transformed into
the following objective
\begin{equation}
\begin{aligned}
\max_{\mathbf{W}_i} \ \  J_i  &= \left\|  \mathbb{Z}^{i} {(\mathbb{Z}^{i})}^{+} {\mathcal{U}}^{i}_s \mathbf{W}_i \right\|_{F}^{2} \\
&=  
         \operatorname{tr}\left\{ \mathbf{W}_i^{\intercal}  (\mathcal{U}_s^i)^{\intercal} \mathbb{Z}^{i} {(\mathbb{Z}^{i})}^{+} {\mathcal{U}}^{i}_s \mathbf{W}_i \right\}        \\ 
    \texttt {s.t. }  \quad &{(\mathbf{V}^{i}_s)}^{\intercal}         \mathbf{V}^{i}_s  = \mathbf{W}_i^{\intercal}    \mathbf{W}_i =\mathbf{I}_{\ell_{i}}.
\end{aligned}
\end{equation}
It is clear that the above objective with $\mathbf{W}_i^{\intercal}    \mathbf{W}_i =\mathbf{I}_{\ell_{i}}$ is achieved only if $\mathbf{W}_i$ are the eigenvectors of  $ (\mathcal{U}_s^i)^{\intercal} \mathbb{Z}^{i} {(\mathbb{Z}^{i})}^{+} {\mathcal{U}}^{i}_s$ corresponding to the $\ell_{i}$ maximum eigenvalues. 
Therefore, performing eigen-decomposition 
\begin{equation}
    (\mathcal{U}_s^i)^{\intercal} \mathbb{Z}^{i} {(\mathbb{Z}^{i})}^{+} {\mathcal{U}}^{i}_s = \mathbf{\bar W}_{i} \bm{\Lambda}_{i} \mathbf{\bar W}_{i}^{\intercal},
\end{equation} 
the optimal solution is $\mathbf{W}_i = \mathbf{\bar  W}_{i} \left(:,1:\ell_{i} \right)$ for $i\in \enumbracket{M}$. The complete iterative Net-LaVARX algorithm is summarized in Algorithm~\ref{alg:netLaVARX}. Note that in the algorithm, the DLV scores are scaled to unit variance as a post-processing step. The loadings matrix is solved by
 \begin{equation}
 \begin{split}
\mathbf{P}_{i} &=\arg \min \|
\mathbf{Y}_s^i - \mathbf{V}_s^i \mathbf{P}_i^\intercal \|_F^2 \\%=(\mathbf{Y}_s^i)^\intercal\mathbf{V}_s^i ((\mathbf{V}_s^i)^\intercal\mathbf{V}_s^i)^{-1}\\
&= (\mathbf{Y}_s^i)^\intercal\mathbf{V}_s^i 
= \mathbf{\cal  V}^{i}_s \mathbf{\cal D}^{i}_s \mathbf{W}^{i}.    
\end{split}
 \end{equation}

\begin{algorithm}[htbp] 
	\renewcommand{\algorithmicrequire}{\textbf{Input:}}
	\renewcommand{\algorithmicensure}{\textbf{Output:}}
	\caption{Pseudo-code for {Net-LaVARX-CCA}} 
	\label{alg:netLaVARX}  
	\begin{algorithmic}%[1] 
	\Require Scale  output $\mathbf{Y}^{i}$ and inputs $\mathbf{U}^{i}$ to zero mean and unit variance; specify the number of DLVs $\ell_i$ and order ${s}_{i}$ for $i \in \enumbracket{M}$; 
    % input order vector $\mathbf{f}$ with each element $f_{i}$; 
	\Ensure Weight matrices $\mathbf{R}_{i}$, loadings matrices $\mathbf{P}_{i}$, and VARX coefficients $\mathbb{Q}^{i}$, for $i \in \enumbracket{M}$; 
        \State Perform  SVD: $\mathbf{Y}_{s}^{i}  = \mathbf{\cal U}_{s}^{i} \mathbf{\cal D}_{s}^{i} {(\mathbf{\cal V}_{s}^{i})}^{\intercal}$, $i \in \enumbracket{M}$; 
        % \JQ{Be careful. This SVD does not guarantee that $\mathbf{\cal U}_{s}^{i}$ is orthogonal. This is the reason why we need to do SVD on $\mathbf{Y}_{s}^{i}$! } \rev{-Yes, sir}
    	\State Initialize $\mathbf{W}^{i} = \mathbf{I}_{r_i}(:,1:\ell_{i})$, where $r_i$ is the rank of $\mathbf{\cal D}^{i}_s$; % \rev{!!!Take care of the dimension of~$\mathbf{\cal U}_{}^{i}$ and~$\mathbf{W}^{i}$!!! -Yes, we need to align the dimension in coding}
            % \State Initialize $\mathbf{V}^{i}_s = \mathbf{\cal U}^{i}_s \mathbf{W}^{i}$, for $i \in \enumbracket{M}$;
        \State Form $\mathbb{U}^{i}_s =\left[\begin{array}{llll}
        \mathbf{U}^{i}_{s_{}-1} & \mathbf{U}^{i}_{s_{}-2} & \cdots & \mathbf{U}^{i}_{s_{}-s_{i}}
        \end{array}\right]$, $i \in \enumbracket{M}$; % 
            \While {$not \ converge \ \& \ {iter} < {max}\_{iter}$}
                \For{${i=1:M}$}
                    \State Update $i$-th DLV node $\mathbf{V}_{s}^{i} = \mathbf{\cal U}_{s}^{i} \mathbf{W}^{i}$;
                    \State Update the corresponding  $\mathbb{V}^{i}_s$ with $\mathbf{V}_{s}^{i}$; 
                    \State Concatenate augmented DLVs and inputs for $\mathbb{Z}^{i}_{s}$;
                    \State Perform eigen-decomposition on ${({\cal{U}}^{i}_s)}^{\intercal} \mathbb{Z}^{i}_{s} {(\mathbb{Z}^{i}_{s})}^{+} {\cal{U}}^{i}_s $ \\ 
                    \quad \quad \quad \quad \ \ and store the $\ell_i$ leading eigenvectors in $\mathbf{W}^{i}$; 
                %     \State Update $\mathbf{W}^{i} = \mathbf{W}^{i} \left(:,1:\ell_{i} \right)$; 
            \EndFor
            \State Convergence detection on $\mathbf{W}^{i}$;
            \EndWhile 
        \State Calculate $\mathbf{R}_{i} = \sqrt{N-1} \mathbf{\cal V}_{s}^{i} {(\mathbf{\cal D}_{s}^{i})}^{-1} \mathbf{W}^{i}$;
        \State Calculate DLV scores $\mathbf{V}^{i} = \mathbf{Y}^{i} \mathbf{R}^{i}$;
        \State Calculate loading matrix $\mathbf{P}_{i} = \mathbf{\cal  V}_{s}^{i} \mathbf{\cal D}_{s}^{i} \mathbf{W}^{i}/\sqrt{N-1}$;
        \State Calculate VARX coefficients $\mathbb{Q}^{i}$, for $i \in \enumbracket{M}$.
	\end{algorithmic} 
\end{algorithm}

\section{Case Study}

In this section, we present an application of the proposed net-LaVARX model on the Tennessee Eastman process (TEP)~\cite{downs:vogel:1993} with specially simulated dynamic data under decentralized control~\cite{Ricker:1996}. 
The goal is to demonstrate the effectiveness of the dynamic latent networked model in predicting the behavior of the TEP systems and to compare the variable partition strategies with different network partitioning of the process.

The current simulation represents an improved version of the original TEP problem \cite{Bathelt:etal:2015RickerTEP}. We record data from 12 manipulated variables and 22 measured process variables. Unlike the heavily utilized simulated data with a 3-minute interval for steady-state modeling \cite{chiang:russell:braatz:2000}, the dataset used in this study is sampled every 36 seconds to reflect latent dynamics in the data. Nine manipulated variables, namely XMV(1)-XMV(8) and XMV(10)-XMV(11), are treated as the exogenous inputs, while XMEAS(1)-XMEAS(22) are chosen as the process measurements.

The central units in TEP consist of  a reactor, a stripper, and a vapor-liquid separator (Separ.) column with a condenser (Cond.) and a compressor (Comp.). 
By examining the process flow chart in \cite{downs:vogel:1993} and decentralized control in \cite{Ricker:1996},  we respectively model the TEP process as networked dynamic systems with two partition scenarios, which are in Table~\ref{tab:TEP_partition}. 
Partition-I includes the node of the reactor with the condenser, the node of the separator with the compressor, and the node of the stripper.
Partition-II moves the compressor to the node of the separator. 
For comparison, we also tested a model that treats all measurements and inputs as a whole, i.e., the LaVARX model, without considering the networked structure, denoted as \textit{Monolithic}.

\begin{table*}[htbp]
  \centering
  \caption{The TEP variable partition description with different strategies.}
    \begin{tabular}{ccll}
    \toprule
    Partition & Nodes w/ Units & Process Measurements & Exogenous Inputs \\
    \midrule
    \multirow{3}[2]{*}{I} & Reactor\&Cond. & XMEAS(1)-XMEAS(3), XMEAS(6)-XMEAS(9), XMEAS(21) & XMV(1)-XMV(3), XMV(10), XMV(11) \\
          & Separator\&Comp. & XMEAS(5), XMEAS(10)-XMEAS(14), XMEAS(20), XMEAS(22) & XMV(5)-XMV(7) \\
          & Stripper & XMEAS(4), XMEAS(15)-XMEAS(19)  & XMV(4), XMV(8) \\
    \midrule
    \multirow{3}[2]{*}{II} & Reactor & XMEAS(1)-XMEAS(3), XMEAS(6)-XMEAS(9), XMEAS(21) & XMV(1)-XMV(3), XMV(10) \\
          & {Separ.\&Cond.\&Comp.} & XMEAS(5), XMEAS(10)-XMEAS(14), XMEAS(20), XMEAS(22) & XMV(5)-XMV(7), XMV(11) \\
          & Stripper & XMEAS(4), XMEAS(15)-XMEAS(19)  & XMV(4), XMV(8) \\
    % \midrule
    % \multirow{4}[2]{*}{III} & Reactor & XMEAS(1)-XMEAS(3), XMEAS(6)-XMEAS(9), XMEAS(21) & XMV(1)-XMV(3), XMV(10) \\
    %       & {Cond.\&Comp.} & XMEAS(5), XMEAS(20) & XMV(5), XMV(11) \\
    %       & Separator & XMEAS(10)-XMEAS(14), XMEAS(22) & XMV(6), XMV(7) \\
    %       & Stripper & XMEAS(4), XMEAS(15)-XMEAS(19)  & XMV(4), XMV(8) \\
    \bottomrule
    \end{tabular}
  \label{tab:TEP_partition}
\end{table*}

The Net-LaVARX models are built on each partition of the TEP process. For the Net-LaVARX model, the hyper-parameters include the number of DLVs $\ell_i$ and the  AR orders $s_i$.
To select the optimal number of hyper-parameters, the grid search method is utilized, where the first 60\% of the samples are used to train the model, and the next 15\% is used for validation purposes. The last 25\% of the data is reserved for testing the model generalization performance. 
The performance metrics adopted are the coefficient of determination ($\text{R}^2$), the correlations (Corr) coefficients, the root mean squared error (RMSE), and the mean absolute error (MAE) between the predicted and actual measurement values. The optimal number of hyper-parameters $\ell_i^{*}$ and $s_i^*$ with performance metrics on the validation set are depicted in Table~\ref{tab:trn_valid}. It is noticed that the sums of the optimal number of $\ell^{*}$ and $s^*$ for Partition-I and Partition-II are comparable to that of the Monolithic (LaVARX) model, but their prediction performance is rather different. Precisely, the Net-LaVARX models with Partition-I and Partition-II attain the respective best RMSE and MAE, and the best $\text{R}^2$ and Corr on the validation set. 
It also verifies the feasibility of dividing the TEP into a networked system model with three nodes according to its central columns, and the effectiveness in improving the prediction performance.

\begin{table}[t]
  \centering
  \caption{The optimal hyper-parameters according to the performance metrics on the validation set. }
  \setlength{\tabcolsep}{4.75mm}{
    \begin{tabular}{cccc}
    \toprule
          & Monolithic & Partition-I & Partition-II \\
    \midrule
    $\ell^*$ & 19    & [7, 6, 5] & [7, 7, 5]  \\
    $s^*$    & 17    & [6, 6, 6] & [6, 6, 6] \\
    $\text{R}^2$    & 0.919 & \textbf{0.972} & \textbf{0.972}  \\
    $\text{Corr}$  & 0.946 & 0.972 & \textbf{0.983} \\
    $\text{RMSE}$  & 0.232 & \textbf{0.114} & 0.117  \\
    $\text{MAE}$   & 0.186 & \textbf{0.090} & 0.092  \\
    \bottomrule
    \end{tabular}}
  \label{tab:trn_valid}
\end{table}

Once the optimal $\ell_i^{*}$ and $s_i^*$ are decided, we train the final Net-LaVARX models again using the entire training set (first 75\% of the samples). Then, we evaluate the performance of the final model on the remaining 25\% of the samples, with the same metrics given in Table~\ref{tab:test_metrics}. The results show that networked systems with both partitions outperform the performance of the monolithic model, as evidenced in Table~\ref{tab:trn_valid} and Table~\ref{tab:test_metrics}.

\begin{figure}[htbp]
    \centering
    \includegraphics[width=0.825 \columnwidth]{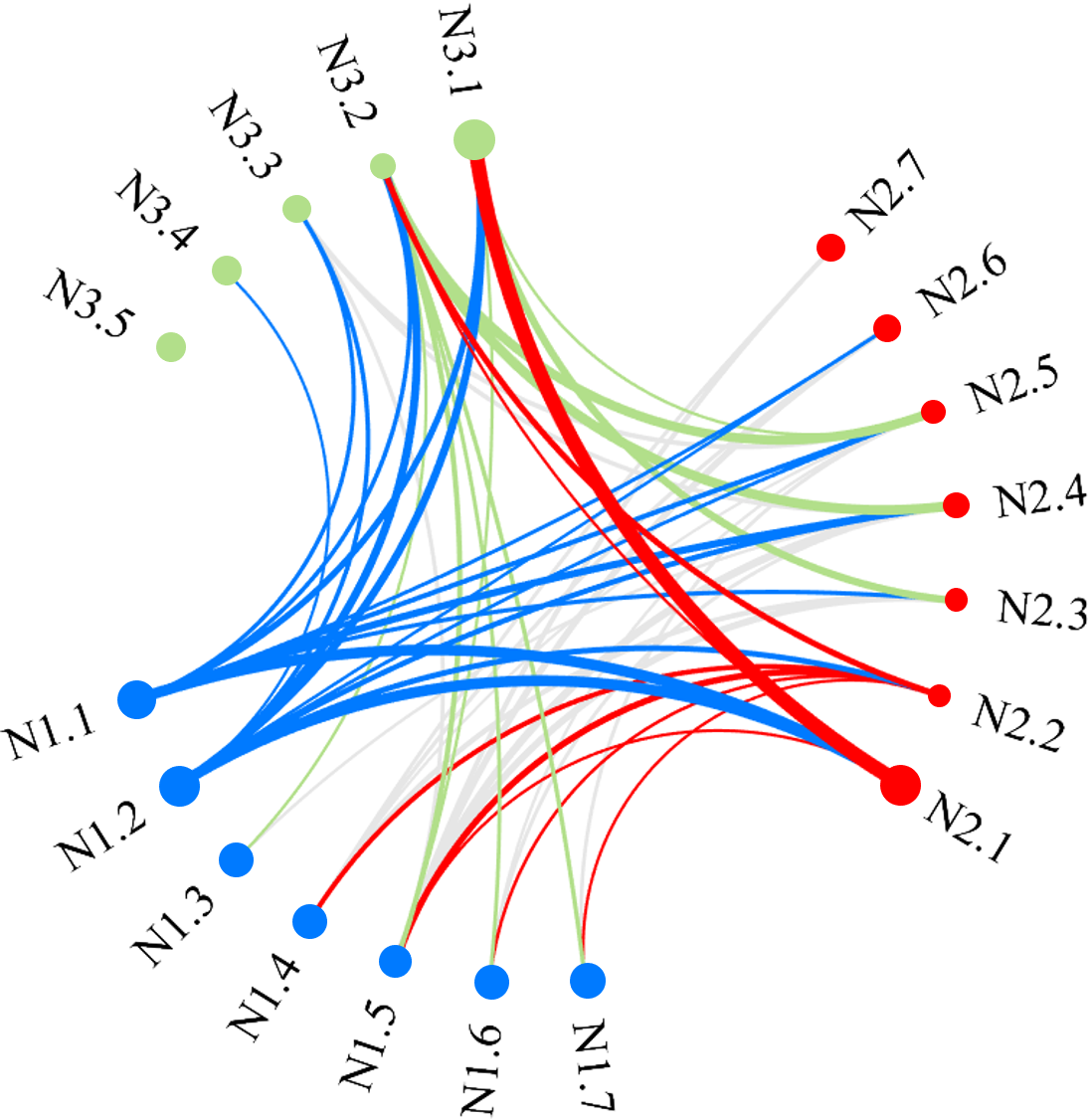}
    \caption{The graph visualization of the networked system nodes according to cross-correlations among the DLV scores in Partition-II.  \text{N1.x}  marked in blue is the reactor node;  \text{N2.x}  in red is the separator node;  \text{N3.x} in green is  the stripper node.
    }
    % \vspace{-1mm}
    \label{fig:netDLV_corr}
\end{figure}

To visualize the interactions between the latent nodes in these networked dynamic systems, the cross-correlations among the DLV scores of each node are calculated. With the threshold of $0.1$ on the correlations, the DLVs network graph for Partition-II is generated and visualized in Fig.~\ref{fig:netDLV_corr}, where the nodes are marked with distinct colors. Each node contains the optimal number of DLVs. 
For easy visualization, the arcs from the leading two DLVs are colored, whereas the remaining ones are shown in light grey. The arcs indicate the  connectivity between these DLVs, and the line width of each arc corresponds to the correlation magnitude between the connected DLVs. It is noticed that the thick arcs of different colors only appear between the leading 2 DLVs among these three nodes. This phenomenon coincides with our model objective that the  predictable DLVs in each node are ranked in descending order. The leading DLVs of all nodes have consistently strong interactions among them, which dominate  the underlying dynamics of the networked system. The minor DLVs have limited predictability and low correlations among them. As a particular case, DLV \text{N3.5} in Fig.~\ref{fig:netDLV_corr} of the stripper node has no significant correlations with other nodes.

In addition, the sizes of dots for  the  node and  DLVs in Fig.~\ref{fig:netDLV_corr} represent the corresponding $\text{R}^2$ values, which correspond to predictability.
It can be seen that the leading DLVs tend to have larger sizes of dots, but this is not absolute. For instance, although \text{N2.2} has a weak intra-group prediction ability, it has considerable connectivity with \text{N1}.
The above observations shed light on interpreting latent dynamic  networked systems for complex plants. 

\begin{table}[t]
  \centering
  \caption{The performance metrics with the optimal hyper-parameters on the test data.}
  \setlength{\tabcolsep}{4.7mm}{
    \begin{tabular}{cccc}
    \toprule
    Metrics & Monolithic & Partition-I & Partition-II  \\
    \midrule
    $\text{R}^2$    & 0.896 & 0.958 & \textbf{0.962}  \\
    $\text{Corr}$  & 0.940 & 0.969 & \textbf{0.979}  \\
    $\text{RMSE}$  & 0.234 & 0.114 & \textbf{0.113}  \\
    $\text{MAE}$   & 0.188 & 0.091 & \textbf{0.089}  \\
    \bottomrule
    \end{tabular}}
  \label{tab:test_metrics}
\end{table}

\section{Conclusions}
In this paper, a new framework for latent dynamic networked system modeling is successfully developed to extract low-dimensional network dynamics from  high-dimensional networked data. The new latent  networked models can handle  co-moving or collinear dynamics with dimension reduction and take the process network topology into account. The Net-LaVARX models applied to the dynamic TEP dataset show that it is beneficial to partition complex plants into networked systems by employing the process topology.  
The results demonstrate the potential of the proposed latent dynamic  networked modeling for system identification.
Future work will focus on the computational complexity and robustness analysis of the networked system identification framework.

\section*{Appendix}

\subsection{Proof of Lemma~\ref{lem:proj_matrix}}
We first need to show that $(\mathbf{\bar  P}_i^\perp)^\intercal\mathbf{  P}_i$ is invertible. Since  
$[\mathbf{  P}_i \quad  \mathbf{\bar  P}_i ] $ in \eqref{eq:NLaVARX_outer}
is non-singular, 
\[
(\mathbf{\bar  P}_i^\perp)^\intercal [\mathbf{  P}_i \quad  \mathbf{\bar  P}_i ] =
[ (\mathbf{\bar  P}_i^\perp)^\intercal\mathbf{  P}_i \quad  \mathbf{0}_{\ell_i \times (p_i -\ell_i)} ] 
\]
has a rank of $\ell_i$, i.e., $(\mathbf{\bar  P}_i^\perp)^\intercal\mathbf{  P}_i$ has full rank. Therefore, $(\mathbf{\bar  P}_i^\perp)^\intercal\mathbf{  P}_i$ is invertible.

It is straightforward to show  that~$$\mathbf{R}_i^\intercal \mathbf{P}_i = ((\mathbf{\bar  P}_i^\perp)^\intercal\mathbf{  P}_i )^{-1}(\mathbf{\bar  P}_i^\perp)^\intercal\mathbf{  P}_i =\mathbf{I}.$$ It follows that
$$(\mathbf{P}_i \mathbf{R}_i ^\intercal)^2 = \mathbf{P}_i (\mathbf{R}_i ^\intercal\mathbf{P}_i) \mathbf{R}_i ^\intercal=
\mathbf{P}_i \mathbf{R}_i ^\intercal, $$
which shows that 
$\mathbf{P}_i \mathbf{R}_i ^\intercal$ is an oblique projection \cite{banerjee:Roy:2014Linear}. 

Moreover, it follows from~$\mathbf{\bar  P}_i^\intercal \mathbf{\bar  P}_i^\perp =\mathbf{0}$ that~$$\mathbf{R}_i^\intercal \mathbf{\bar P}_i= ((\mathbf{\bar  P}_i^\perp)^\intercal\mathbf{  P}_i )^{-1}(\mathbf{\bar  P}_i^\perp)^\intercal\mathbf{\bar P}_i=\mathbf{0}.$$
Thus, we have~
$$\mathbf{R}_i^{\intercal} \bm y^i_k=\mathbf{R}_i^{\intercal}\mathbf{P}_i \bm v^i_k
    + \mathbf{R}_i^{\intercal} \mathbf{\bar P}_i \bm{\bar \varepsilon} ^i_k=\bm v^i_k+\mathbf{0}=\bm v^i_k. $$
This completes the proof.
    
%To show~$\mathbf{P}_i \mathbf{R}_i ^\intercal$ is an oblique projection matrix onto~$\mathbf{P}$ along~$\mathbf{\bar P}$, it remains to show that for an arbitrary~$\bm x \in \Re^p$, the vector~$\mathbf{P}_i \mathbf{R}_i ^\intercal\bm x$ belongs to the subspace spanned by~$\mathbf{P}_i$ and~$\bm x-\mathbf{P}_i \mathbf{R}_i ^\intercal\bm x$ belongs to the subspace spanned by~$\mathbf{\bar P}_i$. The former is trivial and we prove the latter by~$\mathbf{R}_i^\intercal \mathbf{\bar P}_i==\mathbf{0}$ and~$$\mathbf{R}_i^{\intercal}(\bm x-\mathbf{P}_i \mathbf{R}_i ^\intercal\bm x)=\mathbf{R}_i^{\intercal}\bm x-\mathbf{R}_i^{\intercal}\mathbf{P}_i \mathbf{R}_i ^\intercal\bm x=\mathbf{R}_i^{\intercal}\bm x-\mathbf{R}_i^{\intercal}\bm x=\mathbf{0},$$

\balance

\bibliographystyle{IEEEconf} 

\bibliography{IEEEabrv, abbrev, articles, proceedings, books, DLV_Network_Modeling}

\end{document}